\documentstyle[twocolumn,prb,aps,epsfig]{revtex}

\begin{document}

\title{Stationary solutions of the one-dimensional nonlinear Schr\"odinger equation: II. Case of attractive nonlinearity}

\author{L.D. Carr$^{1}$\cite{byline}, Charles W. Clark$^{2}$, W.P. Reinhardt$^{1,2,3}$\\}

\address{$^{1}$Department of Physics, University of Washington, Seattle, WA 98195-1560, USA\\}

\address{$^{2}$Electron and Optical Physics Division, National Institute of Standards and Technology, Technology Administration, US Department of Commerce, Gaithersburg, MD 20899, USA\\}

\address{$^{3}$Department of Chemistry, University of Washington, Seattle, WA 98195-1700, USA\\}

\date{\today}

\maketitle

\begin{abstract}

All stationary solutions to the one-dimensional nonlinear Schr\"odinger equation under box or periodic boundary conditions are presented in analytic form for the case of attractive nonlinearity.  A companion paper has treated the repulsive case.  Our solutions take the form of bounded, quantized, stationary trains of bright solitons.  Among them are two uniquely nonlinear classes of nodeless solutions, whose properties and physical meaning are discussed in detail.  The full set of symmetry-breaking stationary states are described by the $C_{n}$ character tables from the theory of point groups.  We make experimental predictions for the Bose-Einstein condensate and show that, though these are the analog of some of the simplest problems in linear quantum mechanics, nonlinearity introduces new and surprising phenomena.

\end{abstract}



\narrowtext 



\section{Introduction}
\label{sec:intro}
In a recent experiment the dilute-gas, attractive, Bose-Einstein condensate (BEC) was created for the first time for lithium\cite{bradley1,bradley2,sackett1}.  As predicted by the nonlinear Schr\"odinger\cite{gross1,pitaevskii1} equation (NLSE) for three dimensions\cite{dodd1}, the condensate collapsed when the number of particles became large.  However, in one or quasi-one dimension no collapse is predicted~\cite{carr22}, and in one dimension the NLSE has wide application in fiber optics\cite{hasegawa1} as well as other fields\cite{hasimoto1,ciao1,mamaev1,sulem1}.

In this article we present stationary solutions to the one-dimensional NLSE for attractive nonlinearity under periodic and box boundary conditions.  In the preceeding article\cite{carr15} we solved the analogous problem for repulsive nonlinearity.  Although we will place our results in the context of attractive atomic interactions in the BEC, they are equally applicable to wave phenomena in many physical situations, as for example ring lasers\cite{arbel1}.

We define the BEC to be in the quasi-one-dimensional regime when its transverse dimensions are on the order of its healing length and its longitudinal dimension is much longer than its transverse ones.  In this case the 1D limit of the 3D NLSE is appropriate, rather than a true 1D mean field theory~\cite{kolomeisky1}, as would be the case for a transverse dimension on the order of the atomic interaction length.  Under these criteria the condensate is well out of the Thomas-Fermi limit; i.e. the kinetic energy in the transverse dimensions is very high.  It is this high kinetic energy which prevents the condensate from collapsing.  We have numerically illustrated the stability of the condensate in quasi-1D elsewhere\cite{carr22}.

As was shown in the previous article\cite{carr15}, in the quasi-1D regime the solutions are approximately separable, and the dimensionless NLSE may be written:

\begin{equation}
[-\lambda^2\partial_{\tilde{x}}^{2}-\mid\!f(\tilde{x})\!\mid^{2}+\tilde{V}(\tilde{x})\,]\,f(\tilde{x}) = \tilde{\mu}\,f(\tilde{x})
\label{eqn:nlse}
\end{equation}
where $f(\tilde{x})$ is the dimensionless wavefunction describing excitations along $L$, $\tilde{x}\in[0,1]$, $\tilde{V}(\tilde{x})$ is a dimensionless potential representing the boundary conditions, $\tilde{\mu}\equiv (2m\xi^{2}/\hbar^{2})\mu$, and $\lambda\equiv\xi/L\propto N^{-1/2}$, the ratio of the healing length to the box/ring length.  We remind the reader that $\xi^2\equiv 1/(8\pi\bar{\rho}\mid\!a\!\mid)$, where $a$ is the scattering length and $\bar{\rho}$ is the mean particle density. 

For comparison with experiment we list the conversion factor from the unitless $\tilde{\mu}$ to $\mu$ in $\mu$K.  Using the formula from the previous article\cite{carr15}, for $^7$Li with $\bar{\rho}=10^{14}\text{cm}^{-3}$ and $a=-1.45$ nm, the conversion factor is 0.125.  Since the unitless chemical potentials we find will be on the order of -1 to -10 this gives a sense of the energy scale of the solutions, on the order of 0.1 to 1.0 $\mu$K.  We also note that throughout our presentation we will use a reasonable test scale of $\lambda=1/25$ for illustrative purposes.

As $\mid\!f(\tilde{x})\!\mid^{2}$ is the longitudinal portion of the single particle density, we require the normalization condition:

\begin{equation}
\int_{0}^{1} d\tilde{x} \mid\!f(\tilde{x})\!\mid^2 = 1
\label{eqn:normCond}
\end{equation}
be satisfied together with the NLSE (\ref{eqn:nlse}) and such boundary conditions as we will describe below.  These are the equations we will solve.

\section{Box Boundary Conditions}
\label{sec:bbc}
The key concept underpinning the approach taken in this paper is that in a region of constant potential the NLSE is integrable.  This was utilized in our companion paper\cite{carr15}, and virtually all the techniques applied there to the case of repulsive nonlinearity apply here as well, though they lead to the identification of different particular solutions.

Specifically, it was shown in the previous article\cite{carr15} that solutions of the stationary NLSE on the unit interval, subject to the box boundary conditions

\begin{equation}
f(0)=f(1)=0
\label{eqn:boxBC}
\end{equation}
can be expressed, without loss of generality, by purely real functions.  Such solutions are given in terms of Jacobian elliptic functions, whose properties were reviewed in the previous article.  This result holds independent of the sign of the nonlinearity.  For the case of repulsive nonlinearity, it was established that the sn function provides all possible solutions.  For the attractive case, both the cn and dn functions offer solutions.  However, the dn function has no real zeros, so the cn function provides all solutions which satisfy box boundary conditions.

\subsection{Solutions and spectra}

The most general form of the solution is:

\begin{equation}
f(\tilde{x})=A\, \text{cn}(k \tilde{x}+\delta\mid m)
\label{eqn:boxAnsatz}
\end{equation}
where $0\leq m\leq 1$ is the parameter of Jacobian elliptic functions.  $k$ and $\delta$ will be determined by the boundary conditions below while $A$ and $m$ will be determined by substitution of Eq.~(\ref{eqn:boxAnsatz}) into the NLSE and by normalization.

Since the quarter period of cn is the complete elliptic integral of the first kind, $K(m)$, and since cn$(0\mid m)=1$, we find that $k=2jK(m)$ and $\delta=-K(m)$, where $j\in\{1,2,3,...\}$.  The physical meaning of $j-1$ is the number of nodes in the cn function.  We will give a more general meaning to $j$ below.  We then solve Eq.~(\ref{eqn:nlse}) by substituting Eq.~(\ref{eqn:boxAnsatz}), using Jacobian elliptic identities, and setting coefficients of equal powers of cn equal.  This results in equations for the amplitude squared, $A^2$ and the chemical potential, $\tilde{\mu}$ :

\begin{equation}
A^{2}=2 m (2 j K(m))^{2}\lambda^2
\label{eqn:bbcAmp}
\end{equation}

\begin{equation}
\tilde{\mu}= -(2 j K(m))^{2}\lambda^2(2m-1)
\label{eqn:bbcEnergy}
\end{equation}

Substituting Eq.~(\ref{eqn:bbcAmp}) into Eq.~(\ref{eqn:normCond}), utilizing Jacobian elliptic identities, and noting that the integral over cn$^{2}$ can be defined in multiples of the quarter period $K(m)$, we obtain the normalization condition:

\begin{equation}
2 (2j)^{2} \lambda^2 K(m) (E(m) - (1-m)K(m)) = 1
\label{eqn:bbcNormCond}
\end{equation}
where $E(m)$ is the complete elliptic integral of the second kind.  Eq.~(\ref{eqn:boxAnsatz}) then becomes:

\begin{equation}
f(\tilde{x})=\sqrt{2 m}(2 j K(m))\lambda\,\text{cn}(K(m)(2 j \tilde{x} -1)\mid m)
\label{eqn:bbcWavefn}
\end{equation}

\begin{figure}
\begin{center}
\epsfig{figure=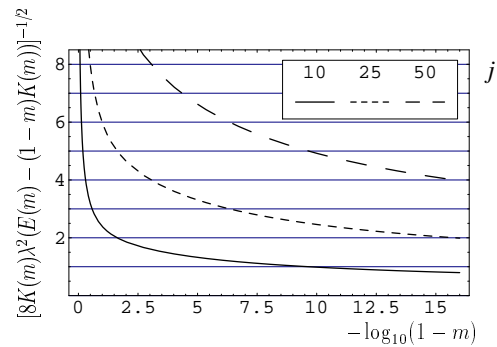,width=8.2cm}
\end{center}
\caption{
This graphical solution of Eq.~(\ref{eqn:bbcNormCond}) shows that for a given scale and number of nodes the real solution to the stationary NLSE under box or periodic boundary conditions is unique.  $\lambda$ is the scale and $j-1$ with $j\in\{1,2,3,...\}$ or $j$ with $j\in\{2,4,6,...\}$ is the number of nodes, respectively.  The three curved lines are plots of Eq.~(\ref{eqn:bbcNormCond}) solved for the number of nodes $j$, with $\lambda^{-1}=L/\xi=10,25,50$.  The left-hand side of the plot is the $m=0$, linear limit, while the right-hand side exponentially approaches the $m=1$, bright soliton limit.  The solutions are found where these lines intersect with the horizontal lines of $j$.  Note the rapid convergence to $m=0$ in the high $j$ limit, so that for large $j$ the solutions are in the linear regime.
}
\label{fig:FnormCond}
\end{figure}

This leaves the chemical potential (\ref{eqn:bbcEnergy}) and the wavefunction (\ref{eqn:bbcWavefn}) determined up to the parameter $m$ and the scale $\lambda$.  In Fig.~\ref{fig:FnormCond} a graphical solution of Eq.~(\ref{eqn:bbcNormCond}) is shown.  The plot demonstrates that the solutions are unique.  Note that the normalization condition requires $m$ to be much closer to 1 than in the repulsive case, as may be seen by comparing Fig.~\ref{fig:FnormCond} to the corresponding figure in the previous article\cite{carr15}.  It follows that the numerics of the attractive case are more difficult than those of the repulsive one, as may be seen in some of the figures.

These solutions are in one-to-one correspondence with those of the 1D particle-in-a-box problem in linear quantum mechanics.  Plots of the wavefunction for the ground state and the first three excited states are shown in Fig.~\ref{fig:FbbcProbAmp}.  In Fig.~\ref{fig:FbbcSpectrum} the chemical potential spectrum of this solution type is plotted as a function of $\lambda^{-2}$, the number of healing lengths per box length quantity squared.  The leftmost portion of the plot corresponds to the particle-in-a-box limit and the rightmost portion to the bright soliton limit.  We now discuss these two limits.

\subsection{Particle-in-a-box limit}

High chemical potential states in which the peaks overlap become particle-in-a-box type solutions, as can be seen in Fig.~\ref{fig:FbbcProbAmp}d.  This is both the zero density, linear limit and the highly-excited-state limit.  Mathematically, $m\rightarrow 0^{+}$ and $\text{cn}\rightarrow \cos$.  Physically, $j\lambda\gg 1$.  In this limit $K(m)\rightarrow \pi(1/2 + m/8 + {\cal O}(m^2))$ and $m\rightarrow 1/(j\pi\lambda)^2$, so that Eq.~(\ref{eqn:bbcEnergy}) becomes:

\begin{eqnarray}
\tilde{\mu}= j^2\pi^2\lambda^2(1-\frac{3m}{2}+{\cal O}(m^2)) \nonumber \\
\tilde{\mu}= j^2\pi^2\lambda^2(1 -\frac{3}{2j^2\pi^2\lambda^2}+{\cal O}(\frac{1}{j^4\lambda^4}))
\end{eqnarray}
or putting back in the units:

\begin{eqnarray}
\label{eqn:perturb}
\mu= \frac{j^{2}\pi^{2}\hbar^{2}}{2 M L^{2}}(1-\frac{3m}{2}+{\cal O}(m^2)) \nonumber \\
\mu= \frac{j^{2}\pi^{2}\hbar^{2}}{2 M L^{2}}(1 -\frac{12 a N L}{A_t j^2\pi}+{\cal O}(\frac{L^2 N^2}{j^4}))
\end{eqnarray}
which clearly converges to the well-known linear quantum mechanics particle-in-a-box chemical potential.  Note that the first order correction is identical to the repulsive case except that it is negative.  One may also obtain this result from first order perturbation theory, as has been shown in the previous article\cite{carr15}.

\subsection{Bright soliton limit}

\begin{figure}
\begin{center}
\epsfig{figure=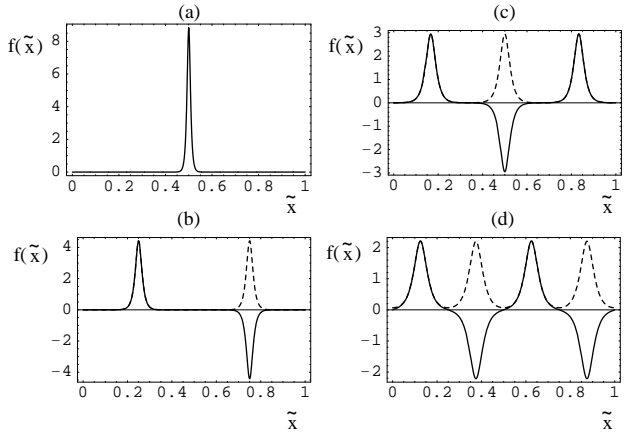,width=8.2cm}
\end{center}
\caption{
Real stationary solutions to the NLSE under box and periodic boundary conditions.  The solid lines show solutions in one-to-one correspondence with those of the analogous particle-in-a-box and particle-on-a-ring problems in linear quantum mechanics.  The dashed lines show uniquely nonlinear, nodeless solutions found only on the ring.  Both solution-types may be characterized as bright soliton trains.  Box: the solid lines in (a)-(d) are the ground state and first three excited states.  Ring: (a) \emph{nodeless, symmetry-breaking} ground state (note that the solid line has overwritten the dashed one) (b)-(d) symmetric and anti-symmetric solutions (c) only the dashed line is a solution.  All plots are for the test scale of $\xi/L=1/25$.
}
\label{fig:FbbcProbAmp}
\end{figure}

One may add a peak without disturbing another peak, provided that adjacent peaks have opposite phase and that the overlap between them is exponentially small in the ratio of their separation to their healing length.  In this limit we ought to recover a series of equally spaced sech solutions of alternating phase to within exponentially small factors, as for example is shown in Fig.~\ref{fig:FbbcProbAmp}b and Fig.~\ref{fig:FbbcProbAmp}c.  Sech solutions are also called bright solitons.  The sech function is the $m\rightarrow 1^{-}$ limit of the cn function in the solution, Eq.~(\ref{eqn:boxAnsatz}).

Bright solitons solve the free NLSE.  Thus we should find that the wavefunction and chemical potential no longer depend on the box length $L$.  In this limit the solitons must have a different length scale than $\lambda$, which depended on $L$.  The soliton width is proportional to the
parameter

\begin{equation}
\eta \equiv \frac{A_t}{8\pi N\mid\!a\!\mid},
\end{equation}
where $A_t$ is the transverse area of the box.  In the soliton literature\cite{hasegawa1}, $\eta$ is usually set to 1 by renormalizing the wavefunction.

We now consider the limit $\lambda\rightarrow 0$, which corresponds to $m\rightarrow 1^{-}$.  Physically, this means that the peaks become highly separated and the interaction between them becomes exponentially small.  By using Taylor expansions in $(1-m)$ of the complete elliptic integrals in Eq.~(\ref{eqn:bbcNormCond}) we find that Eq.~(\ref{eqn:bbcEnergy}) becomes:

\begin{equation}
\tilde{\mu}=-\frac{1}{16j^2\lambda^2}
\end{equation}
while Eq.~(\ref{eqn:bbcWavefn}) becomes:

\begin{equation}
f(\tilde{x})=\frac{1}{2^{3/2}j}\lambda^{-1}\text{cn}(\frac{1}{8j^{2}\lambda^{2}}(2j\tilde{x}-1)\mid m)
\end{equation}
Putting back in the units we find:

\begin{equation}
\mu=-\frac{\hbar^{2}}{2m}\frac{1}{16\eta^2}\frac{1}{j^{2}}
\label{eqn:solitonEnergy}
\end{equation}

\begin{equation}
\frac{f(x)}{\sqrt{L}}=\frac{1}{2^{3/2}j}\frac{1}{\eta}\text{cn}((\frac{1}{4j\eta})x-\delta (L)\mid m)
\end{equation}
where $\delta (L)$ is an offset which depends on the box length.  But as $L\rightarrow \infty$ the offset becomes arbitrary, so that we can set it to zero.  Note that we put back in the units of the wavefunction $f(x)$ which we took out in making the separation of variables in the quasi-1D approximation.

As the reader may verify, for the case in which $j=1$ this is indeed the chemical potential and wavefunction of the sech solution to the free 1D NLSE:

\begin{equation}
\frac{f(x)}{\sqrt{L}}=\frac{1}{2^{3/2}}\frac{1}{\eta}\text{sech}[(\frac{1}{4\eta})x\mid m)
\end{equation}

We have found that this limit suffices to calculate chemical potentials for which $j<1/(5\lambda)$ to better than $1\%$.  This estimate assumes an overall scale size of $\sim 5\xi$ per peak.  The $L$-independent chemical potentials for the $j=1$, $j=2$, and $j=3$ solutions shown in Fig.~\ref{fig:FbbcProbAmp} satisfy this criterion, for example, as do any ground states for a healing length smaller than $1/10$.

Note that Eq.~(\ref{eqn:solitonEnergy}) is identical in form to that of the Rydberg formula for the energy levels of the Hydrogenic atom.  Thus to leading order we have found stationary state energy levels for which the quantum number scales as $j$ (harmonic oscillator)\cite{carr15}, $j^2$ (particle-in-a-box), and $j^{-2}$ (Rydberg formula).  This is an impressive illustration of the complexity of solution types for stationary states of the NLSE.

\section{Periodic boundary conditions}
\label{sec:pbc}

There are four solution types for periodic boundary conditions.  There are constant amplitude solutions which are plane waves; real, antisymmetric, symmetry-breaking solutions, similiar to those found in Sec.~\ref{sec:bbc}; real, symmetric, symmetry-breaking solutions; and a novel class of complex symmetry-breaking solutions.  The former two are in one-to-one correspondence with particle-on-a-ring solutions in linear quantum mechanics; the latter two are \emph{nodeless} and are found only in the presence of nonlinearity.  As the ring is rotationally invariant, the symmetry-breaking solutions will have a high degeneracy, in analogy with vortices in two dimensions\cite{kosterlitz1}.  The periodic boundary conditions are:

\begin{equation}
f(0)=f(1)
\label{eqn:pbcRealBC1}
\end{equation}

\begin{equation}
f^{\prime}(0)=f^{\prime}(1)
\label{eqn:pbcRealBC2}
\end{equation}

\subsection{Constant amplitude solutions}
\label{subsec:cas}

If we assume that $r(\tilde{x})=$ constant then we obtain constant amplitude solutions of the form:

\begin{equation}
f(\tilde{x})=e^{\imath 2\pi n \tilde{x}}
\label{eqn:constAmp}
\end{equation}
where $n \in \{0,\pm 1,\pm 2, ...\}$.  The amplitude is constrained by normalization to be 1.  Substituting Eq.~(\ref{eqn:constAmp}) into Eq.~(\ref{eqn:nlse}) we find the chemical potential:

\begin{equation}
\tilde{\mu}=-1+(\lambda2\pi n)^{2}
\end{equation}

Unlike for the case of repulsive nonlinearity, where the ground state on a ring was the constant solution for $n=0$, the ground state of the attractive BEC breaks symmetry at some scales, so that the constant amplitude solutions are a highly excited state.  This will be discussed further in Sec.~\ref{subsec:gs}.  For $n\neq 0$ each solution is two-fold degenerate, as $n$ can be either positive or negative.

Note that these states could also be termed angular momentum eigenstates or quantized vortices, as for example in the work of Matthews {\it et al.}\cite{matthews1}

\subsection{Real symmetry-breaking solutions}

Of the two Jacobian elliptic functions having different physical form and solving the NLSE for attractive nonlinearity, only one could be used for the box, the cn function.  The dn function satisfied the NLSE but, as it never vanished, failed to satisfy box boundary conditions.  But on the ring the dn function poses no difficulty, because the boundary conditions are periodic.  Thus two real solution types are available.  Looking at the plot of the Jacobian elliptic functions in the appendix to the previous article\cite{carr15}, we observe that the cn function is antisymmetric with respect to translation by the half period $2K(m)$, while the dn function is symmetric under such a translation.

\subsubsection{Analog to box boundary condition solutions}
\label{subsubsec:pbc1}

As we have exchanged the ring for the box, Eq.~(\ref{eqn:boxAnsatz}) is the real solution.  One simply changes $k$ from $2jK(m)$ to $4jK(m)$ in order to satisfy Eqs.~(\ref{eqn:pbcRealBC1}) and (\ref{eqn:pbcRealBC2}), i.e. from multiples of the half period to multiples of the whole period.  The number of nodes will be 2j rather than j-1, where $j\in\{1,2,3,...\}$.  We temporarily keep $\delta$ set to 0.  But note that, unlike for box boundary conditions, under periodic boundary conditions $\delta$ is arbitrary.  

Then all the results from section~\ref{sec:bbc} hold with the new $k$, by letting $j\rightarrow 2j$ in all equations.  The energy and wavefunction are determined uniquely by graphical solution of Fig.~\ref{fig:FnormCond}.  In Fig.~\ref{fig:FbbcProbAmp}b and~\ref{fig:FbbcProbAmp}d we show the first two states.  Both the linear quantum mechanics, particle-on-a-ring limit and the bright soliton limits are reproduced.  In the latter the same kind of non-overlapping criterion applies as before.  

We found real solutions by setting $\delta=0$.  If we instead let $\delta$ vary arbitrarily, we obtain the degeneracy inherent in these symmetry-breaking solutions.  The entropy associated with a pair of peaks depends logarithmically on the box length L, and, since there are approximately $\lambda^{-1}$ possible positions for a peak, the entropy is\cite{carr15,kosterlitz1}:

\begin{equation}
S\sim k_{\text{B}}\text{ln}(\frac{1}{10j\lambda})
\label{eqn:entropy1}
\end{equation}
where the factor of 10 comes from 5 for each of the two peaks.  This is consistent with the non-overlapping criterion used in obtaining Eq.~(\ref{eqn:solitonEnergy}).

\subsubsection{Nodeless solutions}
\label{subsubsec:pbc2}

To find the dn solutions, we follow the same method as outlined in Sec.~\ref{sec:bbc}.  The most general solution is:

\begin{equation}
f(\tilde{x})=A\,\text{dn}(k \tilde{x}+\delta\mid m)
\label{eqn:ringAnsatzDN}
\end{equation}
subject to the NLSE (\ref{eqn:nlse}), the normalization (\ref{eqn:normCond}), and the boundary conditions (\ref{eqn:pbcRealBC1}) and (\ref{eqn:pbcRealBC2}).  From these we obtain the chemical potential, normalization, and amplitude:

\begin{equation}
\tilde{\mu}= -(2 j K(m))^{2}\lambda^2(2-m)
\label{eqn:pbcEnergyDN}
\end{equation}

\begin{equation}
2 (2j)^{2} \lambda^2 K(m) E(m) = 1
\label{eqn:pbcNormCondDN}
\end{equation}

\begin{equation}
f(\tilde{x})=\sqrt{2}(2 j K(m))\lambda \text{dn}(K(m)(2 j \tilde{x} -1)\mid m)
\label{eqn:pbcWavefnDN}
\end{equation}
where $j$ refers to the number of peaks rather than nodes in the dn function, since the dn function is nodeless.  Note that Eqs.~(\ref{eqn:pbcEnergyDN})-(\ref{eqn:pbcWavefnDN}) are only valid for $m\neq 0$.  The $m=0$ case is discussed in Sec.~\ref{subsec:bounds}.

The dn function has a period of $2K(m)$, not $4K(m)$ as did the cn function.  Plots of the amplitude for the first four energy levels are shown in Fig.~\ref{fig:FbbcProbAmp}.  The lowest state, shown in Fig.~\ref{fig:FbbcProbAmp}a, appears identical to the lowest state in Sec.~\ref{sec:bbc}.  Because $m$ is very close to 1 for the test scale of $\lambda=1/25$, dn $\sim$ sech and cn $\sim$ sech, so that cn $\sim$ dn.  Had we chosen a scale at which the peaks came near to the boundaries, the two ground state solutions would have looked quite different.  To make clear the extent to which $m$ is singular at this scale, the numerical solution to Eq.~(\ref{eqn:pbcNormCondDN}) for a single peak is $(1-m)\simeq O(10^{-66})$.

The particle-on-a-ring limit for the dn solution is a plane wave.  As $m\rightarrow 0^{+}$, dn$ \rightarrow 1$.  In this case the amplitude is constrained to be 1 by the normalization and the chemical potential is $\tilde{\mu}=-1$.

In the bright soliton limit Eq.~(\ref{eqn:pbcNormCondDN}) may be expanded in $(1-m)$ to yield the same result for the chemical potential and amplitude as was found for the cn solutions in a box:

\begin{equation}
\mu=-\frac{\hbar^{2}}{2m}\frac{1}{16j^2}\frac{1}{\eta^{2}}
\end{equation}

\begin{equation}
\frac{f(x)}{\sqrt{L}}=\frac{1}{2^{3/2}j}\frac{1}{\eta}\text{dn}((\frac{1}{4j\eta})x-\delta (L)\mid m)
\end{equation}
where the same criterion as was used for the cn solutions in a box may be applied to the validity of the use of the limit here.

As with the cn solutions, we found the nodeless dn solutions by setting $\delta=0$.  If we instead let $\delta$ vary arbitrarily, we obtain the degeneracy inherent in these symmetry-breaking solutions.  The entropy associated with a peak depends logarithmically on the box length L, and, since there are approximately $\lambda^{-1}$ possible positions for a peak, the entropy is:

\begin{equation}
S\sim k_{\text{B}}\text{ln}(\frac{1}{\pi\sqrt{2}j\lambda})
\label{eqn:entropy2}
\end{equation}
where the factor of $\pi\sqrt{2}$ will be explained in Sec.~\ref{subsec:bounds}.

\subsubsection{Energy splittings}
\label{subsubsec:split}

Because the dn solutions are symmetric and the cn solutions are anti-symmetric, we expect that for even numbers of peaks the chemical potentials of the two solution types should be very close.  For odd numbers of peaks there are no cn solutions.

Removing all factors in common to the two chemical potentials we find from Eq.~(\ref{eqn:bbcEnergy}) and Eq.~(\ref{eqn:bbcNormCond}) for the cn solution and from  Eq.~(\ref{eqn:pbcEnergyDN}) and Eq.~(\ref{eqn:pbcNormCondDN}) for the dn solution:

\begin{equation}
\tilde{\mu}_{\text{cn}}(m_{\text{cn}})\propto -\frac{2m_{\text{cn}}-1}{E[m_{\text{cn}}]-(1-m_{\text{cn}})K[m_{\text{cn}}]}
\end{equation}

\begin{equation}
\tilde{\mu}_{\text{dn}}(m_{\text{dn}})\propto -\frac{2-m_{\text{dn}}}{E[m_{\text{dn}}]}
\end{equation}
where $m_{\text{cn}}$ is found by solving Eq.~(\ref{eqn:bbcNormCond}) numerically and $m_{\text{dn}}$ is found by solving Eq.~(\ref{eqn:pbcNormCondDN}) numerically.  This gives the exact percent splitting:

\begin{equation}
\%_{split} = 
100\frac{\tilde{\mu}_{\text{dn}}(m_{\text{dn}})-\tilde{\mu}_{\text{cn}}(m_{\text{cn}})}
     {\frac{1}{2}\mid\!\tilde{\mu}_{\text{dn}}(m_{\text{dn}})+\tilde{\mu}_{\text{cn}}(m_{\text{cn}})\!\mid}
\label{eqn:splitting}
\end{equation}
which falls off exponentially as $m_{cn},m_{dn}\rightarrow 1^{-}$.

In linear quantum mechanics it is known that for the spatial wavefunctions of two particles the antisymmetric superposition is higher in energy than the symmetric superposition.  The exact opposite is true in the NLSE.  Numerical studies of Eq.~(\ref{eqn:splitting}) show that the symmetric, nodeless, dn solutions are \emph{higher} in chemical potential than the antisymmetric cn solutions.  But just as in linear quantum mechanics, the more the wavefunctions overlap the higher the splitting.

\subsection{Complex symmetry-breaking solutions}

Our treatment is identical to that of the repulsive case\cite{carr15}.  By the same arguments used there, all intrinsically complex solutions to Eq.~(\ref{eqn:nlse}) may be written as a sum over standard elliptic integrals by the use of appropriate Cayley transformations\cite{bowman1}.

Writing the wavefunction as:

\begin{equation}
f(\tilde{x})=r(\tilde{x})e^{\imath \phi (\tilde{x})}
\label{eqn:ampPhase}
\end{equation}
one may divide the NLSE into real and imaginary parts:

\begin{equation}
(S^{\prime})^{2}=-2\:[\:\frac{1}{\lambda^2}S^{3}+\frac{2\tilde{\mu}}{\lambda^2}S^{2}-\beta S+2\alpha^{2}\:]
\label{eqn:nlseDensity}
\end{equation}

\begin{equation}
\phi^{\prime}=\frac{\alpha}{S}
\label{eqn:nlsePhase}
\end{equation}
where $\alpha$ and $\beta$ are undetermined constants of integration and $S\equiv r(\tilde{x})^{2}$ is the single particle density $\mid\!\!f(\tilde{x})\!\!\mid^{2}$.  

From substituting Eq.~(\ref{eqn:ampPhase}) into Eqs.~(\ref{eqn:pbcRealBC1})-(\ref{eqn:pbcRealBC2}) and again taking real and imaginary parts, four boundary conditions are obtained, the important one being phase quantization:

\begin{equation}
\phi (1) - \phi (0) = 2\pi n
\label{eqn:pbcComplexBC2}
\end{equation}
where $n$ is an integer which we will call the phase quantum number.

Thus for the complex solutions, Eq.~(\ref{eqn:nlseDensity}) and Eq.~(\ref{eqn:nlsePhase}) replace the NLSE as the equations to solve, together with four boundary conditions, of which phase quantization is the most important, and the normalization, Eq.~(\ref{eqn:normCond}).

In Secs.~\ref{subsubsec:pbc1} and \ref{subsubsec:pbc2} we showed that the real symmetry-breaking solutions have a density proportional to $\text{cn}^{2}$ and $\text{dn}^{2}$, respectively.  The former of these vanishes at $2j$ points around the ring, while the latter is nodeless.  We look for complex solutions which effectively interpolate between these two real solution types.  The motivation for such a solution will become clear in Sec.~\ref{subsec:group}.  Using our physical intuition, we are again able to bypass the use of Cayley transformations, as we did in the repulsive case.

We remind the reader of the Jacobian elliptic function identity:

\begin{equation}
\text{cn}^{2}(\tilde{x}\mid m)=\frac{1}{m}(\text{dn}^{2}(\tilde{x}\mid m)-(1-m))
\end{equation}

We thus generalize the real symmetry-breaking solutions, Eq.~(\ref{eqn:boxAnsatz}) and Eq.~(\ref{eqn:ringAnsatzDN}), as follows:

\begin{equation}
r^{2}(\tilde{x})=A^2(\text{dn}^{2}(k \tilde{x} + \delta \mid m)-\gamma(1-m))
\label{eqn:complexAnsatz0}
\end{equation}
When $\gamma=0$ the dn solution is recovered; when $\gamma=1$ the cn solution is recovered.  As in the repulsive case, we temporarily set $\delta=0$.  $j$ is to be interpreted as the number of peaks in the density $r(\tilde{x})^{2}$.  We will consider the case of general $\delta$, and thus degeneracy, later.  

Using the solution methods as outlined in the repulsive case\cite{carr15}, $k$ is set to $2jK(m)$ in order to satisfy the boundary conditions.  The solution then becomes:

\begin{equation}
r^{2}(\tilde{x})=A^2(\text{dn}^{2}(2jK(m) \tilde{x}\mid m)-\gamma(1-m))
\label{eqn:complexAnsatz}
\end{equation}

The chemical potential and the parameters $\gamma$, $A^2$, and $\alpha$ are then:

\begin{equation}
\tilde{\mu}=-\frac{3}{2}+12j^{2}\lambda^{2}E(m)K(m)-4j^{2}(2-m)\lambda^{2}K(m)^{2}
\label{eqn:complexEnergy}
\end{equation}

\begin{equation}
\gamma = \frac{-j^{-2}\lambda^{-2}+8E(m)K(m)}{8(1-m)K(m)^{2}}
\end{equation}

\begin{equation}
A^2 = 8 j^{2} \lambda^{2} K(m)^{2}
\end{equation}

\begin{eqnarray}
\alpha & = & \frac{1}{\sqrt{2}}[\:\lambda^{-2}((-1+ 8 j^2 \lambda^{2}E(m)K(m) \nonumber \\
       &   & - 8 j^{2} (1-m)\lambda^{2}K(m)^{2}) *( 1 + 8 j^2\lambda^{2}K(m)^{2} \nonumber \\
       &   & + 64 j^4 \lambda^{4} E(m)^{2} K(m)^{2} -16E(m)(j^{2}\lambda^{2}K(m)  \nonumber \\
       &   & + 4 j^{4}\lambda^{4}K(m)^{3}))) \:]^{1/2}
\label{eqn:alpha}
\end{eqnarray}

This leaves the constant of integration $\alpha$ in Eq.~(\ref{eqn:nlsePhase}), the interpolation parameter $\gamma$, the prefactor to the density $A^2$, and the chemical potential $\tilde{\mu}$, determined up to the number of peaks $j$, the scale $\lambda$, and the parameter $m$.  For a given $\lambda$ and $j$ we then numerically integrate Eq.~(\ref{eqn:nlsePhase}) and adjust $m$ until the phase is quantized on the ring, i.e. until $n$ is an integer.

All parameters are monotonic in $m$.  Furthermore $m$ has an extremely limited range.  Because the solution (\ref{eqn:complexAnsatz}) was designed to interpolate between the two real solution types, outside of the small splitting between the value of $m$ for the real cn and dn solutions (see Sec.~\ref{sec:pbc}A3) there are no complex solutions at all.  We find numerically that outside of these splittings $\alpha$ becomes purely imaginary.  This would invalidate the hypothesis that the phase is real.  Thus to find the complex solutions one need only scan the values of $m$ between $m_{\text{cn}}$ and $m_{\text{dn}}$, so that the algorithm is quite straightforward.

If there are an odd number of peaks then there is no cn solution.  In this case we just consider the odd-peaked cn solution for the box, and treat it as a limiting case of the complex solutions.  By symmetry of the ring, $n$ can be either positive or negative, so that each solution is two-fold degenerate, just as we found for the constant amplitude solutions.

\begin{figure}
\begin{center}
\epsfig{figure=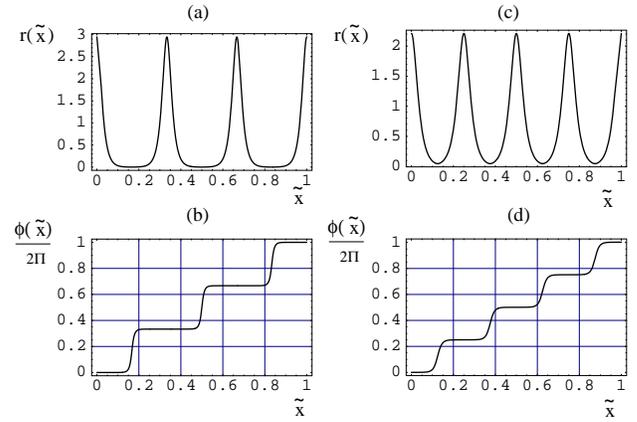,width=8.2cm}
\end{center}
\caption{
Two intrinsically complex, stationary, bright soliton train solutions on a ring.  $j$ is the number of peaks, $n$ is the phase quantum number, and all plots are for the test scale of $\xi/L=1/25$.  (a) Amplitude and (b) phase/$2\pi$ of the $j=3$, $n=1$ solution.  (c) Amplitude and (d) phase/$2\pi$ of the $j=4$, $n=1$ solution.  Note that these solutions are two-fold degenerate, as the chemical potential depends on $n^{2}$.
}
\label{fig:FcomplexAmpPhase1}
\end{figure}

In Fig.~\ref{fig:FcomplexAmpPhase1} we show the lowest two solutions.  We have plotted the amplitude above the phase to make apparent that the phase is nearly constant over the peaks and jumps sharply between them.  When the troughs go to zero, the phase becomes a step function and the height of each step approaches $\pi$, which recovers the cn-type solutions.  As a consistency check, we note that all of the chemical potentials and other parameters in the complex case approached the values found in the real case as $\gamma\rightarrow 0^{+}$ and $\gamma\rightarrow 1^{-}$, respectively.

If $\delta$ is generalized so that it is arbitrary, a similiar degeneracy to what was found in Eqs.~(\ref{eqn:entropy1}) and (\ref{eqn:entropy2}) is obtained:

\begin{equation}
S\sim k_{\text{B}}\text{ln}(\frac{1}{\pi\sqrt{2}j\lambda})
\label{eqn:entropy3}
\end{equation}
where the factor of $\pi\sqrt{2}$ will be explained in Sec.~\ref{subsec:bounds}.

\subsection{Spectra}

\begin{figure}
\begin{center}
\epsfig{figure=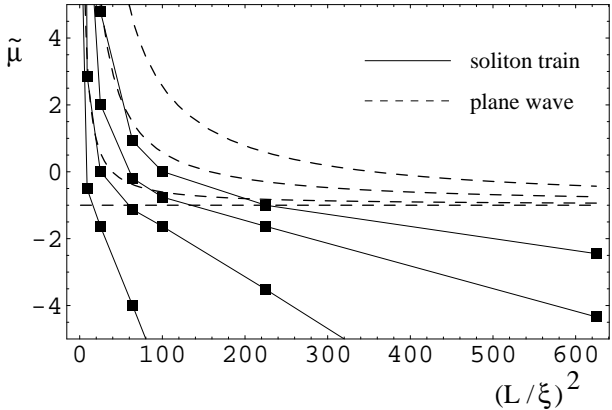,width=8.2cm}
\end{center}
\caption{
Chemical potential spectra of real stationary states with nodes, as a function of inverse scale $L/\xi$, with stationary plane wave spectra shown for comparison.  Dashed lines: shown are $n=0,1,2,3$, where $n$ is the phase quantum-number of the plane wave on the ring.  Solid lines: real stationary states of the NLSE in a box and on a ring are soliton-trains.  Shown are $j=1,2,3,4$ with $j-1$ the number of nodes in a box and $j=2,4$ the number of nodes on a ring.  The linear regime to the far right corresponds to the bright soliton limit in which the peaks are well separated, while the far left corresponds to the particle-in-a-box or particle-on-a-ring limit.
}
\label{fig:FbbcSpectrum}
\end{figure}

\begin{figure}
\begin{center}
\epsfig{figure=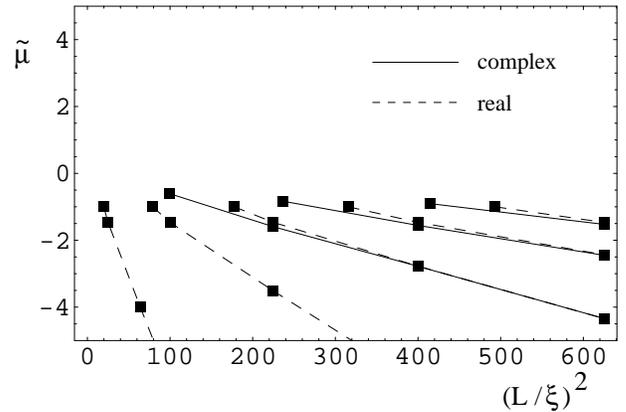,width=8.2cm}
\end{center}
\caption{
Chemical potential spectra for symmetry-breaking, nodeless solutions on the ring.  These solutions have no analogue with the particle-on-a-ring solutions in linear quantum mechanics.  $j$ is the number of peaks, $n$ is the phase quantum number and $L/\xi$ is the number of healing lengths per box length.  Dashed lines: $j=1,2,3,4,5$ from left to right.  Solid lines: $(j,n)=(3,1),(4,1),(5,1)$ from left to right.  The spectra are nearly linear because when the peaks overlap appreciably they no longer solve the nonlinear Schr\"odinger equation, so that they only exist in the bright soliton regime.
}
\label{fig:FpbcSpectrum}
\end{figure}

We show the chemical potential spectra as a function of $\lambda^{-2}$ for the four types of stationary states on the ring: real with nodes, constant amplitude, real without nodes, and intrinsically complex.  In Fig.~\ref{fig:FbbcSpectrum} the two lowest real spectra are shown.  For comparison we have overlayed the four lowest constant amplitude spectra on the same figure.  In Fig.~\ref{fig:FpbcSpectrum} we show the spectra of the nodeless solutions: the three lowest for the complex ones and the five lowest for the real ones.

The spectra in Fig.~\ref{fig:FpbcSpectrum} are very linear because they are in the bright soliton regime.  As soon as the peaks overlap appreciably these two solution types no longer become available.  However, for the cn spectra shown in in Fig.~\ref{fig:FbbcSpectrum} both the bright soliton limit, to the far right, and the particle-on-a-ring limit, to the far left, hold.

For the experimentally reasonable test scale of $\lambda=1/25$ the ground state is the single-peaked, nodeless dn solution.  The first excitation is the two-peaked anti-symmetric cn solution, the second is the two-peaked dn solution, the third is the three-peaked $n=\pm 1$ complex solution, the fourth is the three-peaked dn solution, and so forth.

Since the real solutions are a limiting cases of the complex solutions, the three symmetry-breaking types scale in the same way and their energy levels do not cross, as will be further explained in the following section.  But the constant amplitude solutions depend differently on scale, so their energy levels can cross with those of the other solutions.

\section{Bounds and symmetries on the ring}

The stationary states presented above are the bounded, quantized version of bright soliton-trains.  Solutions on the ring have many special properties and symmetries.  We detail their bounds, degeneracies, and point group symmetry in light of the BEC.  That is, we consider the effect of changing the scale parameter $\lambda$, which corresponds to accretion of atoms into a condensate.  Experimental predictions will be made in Sec.~\ref{sec:predict}.

\subsection{Bounds}
\label{subsec:bounds}

Box-type stationary states can have an arbitrarily large number of nodes.  But the excitation level of nodeless solutions is limited.  We set three bounds on the nodeless stationary states: the maximum chemical potential; the minimum and maximum phase quantum number; and the minimum scale to obtain $j$ notches.  As a consequence of these bounds there are some scales at which no nodeless solutions exist.

Mathematically, the maximum number of peaks that can fit on the ring is obtained from the lower limit on the period of the dn function in the solution (\ref{eqn:complexAnsatz}).  When the peaks overlap too much they are no longer solutions to the NLSE.  The dn function approaches its minimum period of $\pi$ as $m\rightarrow 0^{+}$.  In this limit Eq.~(\ref{eqn:complexEnergy}) becomes:

\begin{equation}
\tilde{\mu}_{max}=-\frac{3}{2}+2j^{2}\lambda^{2}\pi^{2}
\label{eqn:maxEnergy}
\end{equation}

In this same limit the amplitude approaches a constant which the normalization constrains to be 1.  From Eqs.~(\ref{eqn:nlsePhase}) and (\ref{eqn:alpha}) we find a relation between the maximum number of peaks, the phase quantum number, and the scale $\lambda$:

\begin{equation}
\lambda=\frac{1}{\pi \sqrt{2j_{max}^{2}-8n^{2}}}
\label{eqn:boundRelation}
\end{equation}
If $n=0$ the dn solution is recovered.  Substituting Eq.~(\ref{eqn:boundRelation}) into Eq.~(\ref{eqn:maxEnergy}) and setting $n=0$ we find that $\tilde{\mu}_{max}=-1$.  Therefore the chemical potential of the plane wave solution is an upper bound for all nodeless solutions.

We can set a bound on the phase quanta simply from the parameter range mentioned in Sec.~\ref{sec:pbc}C.  Because the complex solutions interpolate between the cn and dn solutions:

\begin{equation}
0 \leq  n < \frac{j}{2}
\end{equation}
Note that $n$ is stricly less than $j/2$.  This is because the phase quanta of the cn solution is actually 0, not $j/2$.  $j/2$ is the limiting case, and $j$ can be odd, even though an odd-peaked cn solution does not solve the NLSE on the ring.

\begin{figure}
\begin{center}
\epsfig{figure=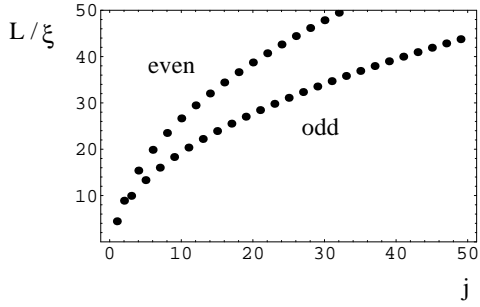,width=8.2cm}
\end{center}
\caption{
Minimum inverse scale for $j$ peaks to become available.  The lower curve is odd $j$; the upper curve is even $j$.  Note that in general there are many more odd solutions than even solutions available.  The ordering of the solutions is $j=(1,2,3,5,4,7,9,6,11,13,...)$.
}
\label{fig:FminScale}
\end{figure}

In Fig.~\ref{fig:FminScale} we plot the inverse scale at which each $j$ becomes available for the maximum $n$ within the range of $n$.  For odd $j$ this is only half a phase quantum away from $j/2$, while for even $j$ it is a whole phase quantum away.  Thus we expect that the odd $j$ solutions should become available at lower inverse scales than the even $j$ solutions, due to the quantization condition:

\begin{eqnarray}
(\lambda^{-1})_{min}=\pi\sqrt{-8+8j_{even}} \nonumber \\
(\lambda^{-1})_{min}=\pi\sqrt{-2+4j_{odd}}
\label{eqn:evenj}
\end{eqnarray}
for even or odd $j$, respectively.  The order of j-peaked solutions turning on as a function of $\lambda^{-1}$ is then:  $1,2,3,5,4,7,9,6,11,13,8,...$  Thus for large $\lambda^{-1}$ many more odd than even solutions are available, as again may be seen in the plot.
  
In Fig.~\ref{fig:FminScale} and in Eq.~(\ref{eqn:evenj}) it is apparent that the minimum inverse scale for a complex solution is $\pi\sqrt{2}$.  This means that the ground state for less than about 4.5 healing lengths to the box length is the $n=0$ constant solution, rather than the single-peaked $n=0$ dn solution.  This is also the source of the constant factor in the entropy of the nodeless solutions, Eqs.~(\ref{eqn:entropy2}) and (\ref{eqn:entropy3}).  $\pi\sqrt{2}$ is the minimum size of a peak.


Supposing that $\lambda^{-1}$ is large there are then three regimes: for small $j$ all solution types are available; for intermediate $j$ even-peaked solutions are cn and go to zero, while odd-peaked solutions are complex and shallow; and for large $j$ only cn solutions are available, so that there are no odd-peaked solutions.



\subsection{Theory of point groups}
\label{subsec:group}

\begin{figure}
\begin{center}
\epsfig{figure=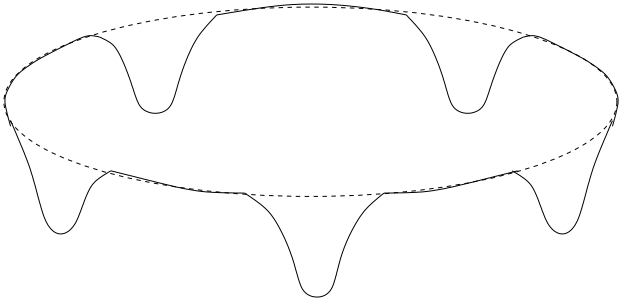,width=8.2cm}
\end{center}
\caption{
Five-fold group symmetry on the ring.  The dips are the mean-field effective potential produced by the five peaks in the condensate.
}
\label{fig:FgroupRing5}
\end{figure}

From Eq.~(\ref{eqn:nlse}) one may consider the negative nonlinear term in the NLSE as an effective potential generated by the wavefunction.  As long as the peaks in the wavefunction are well separated then the self-generated troughs do not interact.  On a ring one may treat the well-separated limit as a rotationally symmetric set of $j$ potential wells, where $j$ is, as before, the number of peaks in the magnitude of the wavefunction.  In Fig.~\ref{fig:FgroupRing5} the case for $j=5$ is shown.

This is in complete analogy to the quantum dynamics of a particle in a j-fold rotationally symmetric potential in two dimensions.  Such a physical situation is described by the $C_{j}$ symmetry point group.  The number of irreducible representations of the $C_{j}$ group determines the degeneracy of the solutions.  Thus one may look up the $C_{j}$ character tables~\cite{tinkham1} in order to find the degeneracy.

Consider the case of $j=5$:  from Eq.~(\ref{eqn:evenj}) it may be seen that there is a real, $n=0$, dn solution and two complex solutions, $n=1$ and $n=2$.  As the two complex solutions are each two-fold degenerate there are five solutions in all.  This is identical to the the $C_{5}$ character table~\cite{tinkham1}:  there are one real solution and two doubly-degenerate, intrinsically complex solutions.  The latter are essentially a linear combination of independent bright solitons with coefficients which are simply the appropriate group characters, as familiar to chemists in molecular orbital theory~\cite{cotton1}.

In the limit that $\lambda\rightarrow 0^{+}$, i.e. that the peaks are well separated, given $j$ peaks there are $j$ nearly degenerate solutions.  If $j$ is even there are 2 real solutions, one symmetric and one antisymmetric, and $(j-2)/2$ complex solutions, each two-fold degenerate.  The splitting is given by Eq.~(\ref{eqn:splitting}).  If $j$ is odd there is only one real solution, the symmetric one, and $(j-1)/2$ complex solutions, each two-fold degenerate.

An upper bound may be placed on the splitting by applying Eq.~(\ref{eqn:splitting}) and keeping in mind that the odd-peaked cn solution is a limiting case on the ring.  For both even and odd $j$ the ordering of the chemical potentials is as detailed in Sec.~\ref{sec:pbc}D, with the anti-symmetric solution being the lowest and the other symmetry-breaking solutions being higher in chemical potential as $n\rightarrow 0^{+}$.  There is an additional degeneracy due to symmetry-breaking, as seen in Eqs.~(\ref{eqn:entropy1}), (\ref{eqn:entropy2}), and (\ref{eqn:entropy3}).

Thus we have shown by group theoretic considerations that our bright stationary state solutions to the NLSE under periodic boundary conditions include all symmetry-breaking eigenstates of evenly spaced peaks identical up to a phase.

\section{Experimental predictions}
\label{sec:predict}

\subsection{Formation of the ground state}
\label{subsec:gs}

The symmetry of the ground state of the NLSE on a ring is scale-dependent.  For $\lambda^{-1} < \sqrt{2}\pi$ the ground state is a constant; for $\lambda^{-1} > \sqrt{2}\pi$ it is single-peaked, symmetry-breaking, and  nodeless.  The width of this peak is about $\sqrt{2}\pi\lambda$.  Recalling the definition of $\lambda$, for fixed L, i.e. a fixed trap size, we may consider the symmetry-breaking as a function of either particle number or scattering length.  The former case corresponds to the experimental situation of adiabatic accretion of particles in quasi-1D.  The latter corresponds to the tuning of a scattering length, by a Feshbach resonance~\cite{kagan2} or other means.

\begin{figure}
\begin{center}
\epsfig{figure=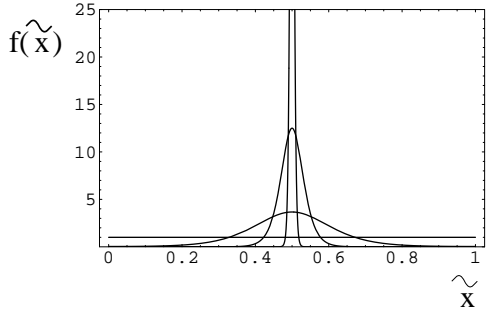,width=8.2cm}
\end{center}
\caption{
Adiabiatic formation of the ground state on a ring: the onset of symmetry-breaking.  From the broadest to the thinnest solution the scales are $L/\xi= \pi\sqrt{2}, 5, 10, 25$.  The respective chemical potentials are $\tilde{\mu}=-1, -1.852,-6.250, -39.025$.  In a box the formation is similiar, except that at the broadest scales the ground state goes smoothly to zero rather than being constant.
}
\label{fig:FadiabatRing}
\end{figure}

Thus one expects under adiabatic growth of the condensate to observe the density of the attractive BEC, as a function of the number of atoms trapped, to be constant, then to peak in the middle, thereby breaking symmetry, and finally to disappear from view as the width of the peak becomes less than the wavelength of the imaging radiation.  In Fig.~\ref{fig:FadiabatRing} we show this sequence of events in four stages superimposed on each other.  Under box boundary conditions one expects a similiar occurrence.  Although the ground state is not a constant, it is quite broad for $\lambda^{-1} \sim 1$.  As $\lambda^{-1}$ increases, the ground state becomes steadily sharper until it is no longer visible under the imaging radiation.

Though it has been shown that the NLSE models the BEC quite well in the repulsive case, it has been unclear as to how well it models the physics of the attractive case.  There have been relatively few experiments with attractive interactions, and analysis of the NLSE in a three-dimensional trapped condensate suggested a small upper limit on the number of atoms that can be maintained in a stable condensate\cite{bradley1,bradley2,sackett1,dodd1}.

We predict that in quasi-1D, i.e. for transverse dimensions on the order of $\xi$, a slow enough accretion of particles to the attractive condensate will have the sequence shown in Fig.~\ref{fig:FadiabatRing} for the ring and a similiar sequence for the box.  An experiment looking for such a change in the shape of the ground state could answer two questions at once: whether there are bright solitons and whether the NLSE is a useful model for the attractive BEC.  If it can be shown that the NLSE models the attractive BEC in quasi-1D then there are a plethora of rich phenomena in fiber optics which could have a direct analogue in the BEC.

\subsection{Modulational instability}
\label{subsec:mod}

For $\lambda^{-1} < \sqrt{2}\pi$ we have shown that the constant solution is the ground state.  We have discussed the case of adiabatic changes in the particle number or the scattering length.  Let us now consider \emph{non-adiabatic} changes.

\begin{figure}
\begin{center}
\epsfig{figure=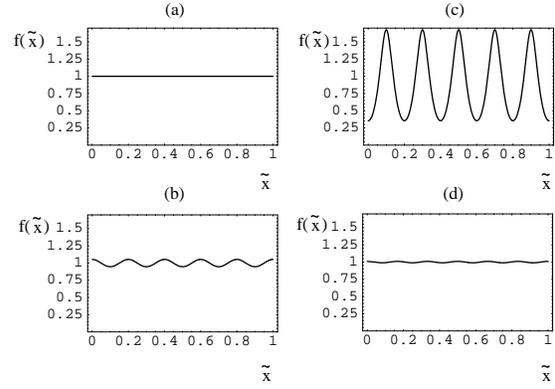,width=8.2cm}
\end{center}
\caption{
Modulational instability: fluctuations in particle number lead to symmetry-breaking solutions of radically varying depth.  All four plots are for $j=5$ peaked dn solutions, with scale varied from the minimal number of healing lengths needed to get a five-peaked solution to just over the minimal number needed to obtain a six-peaked solution.  (a) $L/\xi = 5\pi\sqrt{2}$, $\tilde{\mu}=-1.0000$ (b) $L/\xi = 5\pi\sqrt{2}+10^{-2}$, $\tilde{\mu}=-1.0028$ (c) $L/\xi = 25$, $\tilde{\mu}=-1.4657$ (d) $L/\xi =6\pi\sqrt{2}+10^{-3}$, $\tilde{\mu}=-1.0001$  
}
\label{fig:FmodInstability}
\end{figure}

We first note that the depth of the dn solutions is highly scale dependent.  We solve Eq.~(\ref{eqn:pbcNormCondDN}) in the limit as $m\rightarrow 0^{+}$ for $\lambda^{-1}$:

\begin{equation}
\lambda^{-1}=\sqrt{2}\pi j
\end{equation}
For each $j\in \{1,2,3,...\}$ there is a scale for which the real, j-peaked dn solution is in fact a constant.  Near these scales the dn has shallow modulations.  Midway between these scales it is exponentially close to zero between the peaks.  in Fig.~\ref{fig:FmodInstability} we plot such a sequence for the $j=5$ solution.  The four plots of the wavefunction maintain the same phase quantum number, $n=0$, yet vary greatly in depth with only a small fluctuation of scale.  This extreme sensitivity in the form of the solutions to an external parameter, here $\lambda^{-1}$, is akin to the modulational instability found in fiber optics, in which a continuous wave laser picks up highly variable sidebands in response to small changes in the power~\cite{hasegawa1}.

If the scattering length is positive then the BEC is repulsive and the ground state on a ring is the constant solution.  Using a Feshbach resonance the scattering length may be tuned rapidly negative.  In the case of $^{23}$Na the condensate has been experimentally observed to disappear from view when the scattering length becomes negative~\cite{stenger1,inouye1}, with various explanations based on atomic recombination mechanisms~\cite{timmermans1,abeelen1,yurovsky1}.  In the case of $^{85}$Rb the condensate shrank to below the resolution limit and then emitted a burst of high-energy atoms~\cite{cornish1}.

We have shown here that the stationary NLSE provides a mechanism for the apparent destruction of the condensate even in the absence of recombination, at least in the quasi-1D approximation.  For a fast transition in the scattering length we expect the constant solution, in response to noise, to be able to easily shift to a many-peaked soliton solution without having to change phase quantum number.  It may even be possible to control the form of this transition by using intentional noise to induce the desired j-peaked soliton train solution.  Induced modulation has proven quite useful in fiber optics, as for example in the creation of a new kind of laser~\cite{hasegawa1}.

\section{Conclusion}
\label{sec:conclusion}

We have presented the complete set of stationary solutions to the nonlinear Schr\"odinger equation under periodic and box boundary conditions in one dimension for the case of attractive nonlinearity.  In a box all solutions may be taken to be real.  On a ring there are four solution types: constant amplitude solutions which are plane waves; real symmetry-breaking solutions with nodes; real nodeless symmetry-breaking solutions; and a novel class of complex, symmetry-breaking, nodeless solutions.  All symmetry-breaking stationary states on the ring are described by the theory of point symmetry groups.

Constant amplitude solutions and real solutions with nodes are in one-to-one correspondence with those of the analogous particle-on-a-ring and particle-in-a-box problems in linear quantum mechanics.  Nodeless, symmetry-breaking solutions are uniquely nonlinear.  Solutions of non-constant amplitude may be treated as bright soliton-trains.  As the natural size of a density-peak is $\pi\sqrt{2}$, the minimum scale size needed to obtain nodeless solutions is $L/\xi = \pi\sqrt{2}$.  As a consequence, the form of the ground state depends on $L/\xi$, which in the context of the BEC means that as atoms accrete the ground state on the ring breaks symmetry.

In addition to describing the properties and physical meaning of stationary states in detail, we have made experimental predictions specific to the BEC.  The creation of an attractive BEC in the quasi-one-dimensional regime could answer two questions at once: is the NLSE with attractive nonlinearity a good model; and can one observe solitons?  Thus far experiments have only been performed in the three-dimensional regime.  The quasi-one-dimensional solutions presented in this work may suggest further experiments.  Elsewhere we have illustrated such solutions numerically~\cite{carr22}.


\acknowledgments

We benefited greatly from extensive discussions with Nathan Kutz and David Thouless.  Early stages of this work were supported in part by the Office of Naval Research; the work was completed with the partial support of NSF Chemistry and Physics.



\begin{references}

\bibitem[*]{byline} to whom correspondence should be addressed

\bibitem{bradley1}
C.~C. Bradley, C.~A. Sackett, J.~J. Tollett, and R.~G. Hulet, Phys. Rev. Letts.
  {\bf 75},  1687  (1995).

\bibitem{bradley2}
C.~C. Bradley, C.~A. Sackett, and R.~G. Hulet, Phys. Rev. A {\bf 55},  3951
  (1997).

\bibitem{sackett1}
C.~A. Sackett, J.~M. Gerton, M. Welling, and R.~G. Hulet, Phys. Rev. Lett. {\bf
  82},  876  (1999).

\bibitem{gross1}
E.~P. Gross, Nuovo Cimento {\bf 20},  454  (1961).

\bibitem{pitaevskii1}
L.~P. Pitaevskii, Sov. Phys. JETP {\bf 13},  451  (1961).

\bibitem{dodd1}
R.~J. Dodd {\it et~al.}, Phys. Rev. A {\bf 54},  661  (1996).

\bibitem{carr22}
L.~D. Carr, M.~A. Leung, and W.~P. Reinhardt, J. Phys. B in press, e-print
  cond-mat/0004287 (unpublished).

\bibitem{hasegawa1}
A. Hasegawa, {\em Optical Solitons in Fibers} (Springer-Verlag, New York,
  1990).

\bibitem{hasimoto1}
H. Hasimoto, J. Fluid Mech. {\bf 51},  477  (1972).

\bibitem{ciao1}
R.~Y. Ciao, I.~H. Deutsch, J.~C. Garrison, and E.~W. Wright, {\em Frontiers in
  Nonlinear Optics: the Serge Akhmanov Memorial Volume} (Institute of Physics
  Publishing, Bristol and Philadelphia, 1993), p.\ 151.

\bibitem{mamaev1}
A.~V. Mamaev, M. Saffman, D.~Z. Anderson, and A.~A. Zozuyla, Phys. Rev. A {\bf
  54},  870  (1996).

\bibitem{sulem1}
C. Sulem and P.~L. Sulem, {\em Nonlinear Schr\"odinger Equations: Self-focusing
  Instability and Wave Collapse} (Springer-Verlag, New York, 1999).

\bibitem{carr15}
L.~D. Carr, C.~W. Clark, and W.~P. Reinhardt, submitted to Phys. Rev. A,
  e-print cond-mat/9911177 (unpublished).

\bibitem{arbel1}
D. Arbel and M. Orenstein, IEEE J. Quant. Elect. {\bf 35},  977  (1999).

\bibitem{kolomeisky1}
E.~B. Kolomeisky, T.~J. Newman, J.~P. Straley, and X. Qi, e-print
  cond-mat/0002282 (unpublished).

\bibitem{kosterlitz1}
J.~M. Kosterlitz and D.~J. Thouless, J. Phys. C {\bf 6},  1181  (1973).

\bibitem{matthews1}
M.~R. Matthews {\it et~al.}, Phys. Rev. Letts. {\bf 83},  2498  (1999).

\bibitem{bowman1}
F. Bowman, {\em Introduction to Elliptic Functions, with Applications} (Dover,
  New York, 1961).

\bibitem{tinkham1}
M. Tinkham, {\em Group Theory and Quantum Mechanics} (McGraw-Hill, New York,
  1964).

\bibitem{cotton1}
F.~A. Cotton, {\em Chemical Application of Group Theory}, 3rd ed.
  (Wiley-Interscience, New York, 1960).


\bibitem{kagan2}
Y. Kagan, E.~L. Surkiv, and G.~V. Shylapnikov
Phys. Rev. Letts, {\bf 79}, 2604 (1997).

\bibitem{stenger1}
J. Stenger {\it et~al.}, Phys. Rev. Letts. {\bf 82},  2422  (1999).

\bibitem{inouye1}
S. Inouye {\it et~al.}, Nature {\bf 392},  151  (1998).

\bibitem{timmermans1}
E. Timmermans, P. Tommasini, M. Hussein, and A. Kerman, Phys. Rep. {\bf 315},
  199  (1999).

\bibitem{abeelen1}
F.~A. van Abeelen and B.~J. Verhaar, Phys. Rev. Letts. {\bf 83},  1550  (1999).

\bibitem{yurovsky1}
V.~A. Yurovsky, A. Ben-Reuven, P.~S. Julienne, and C.~J. Williams, Phys. Rev. A
  {\bf 60},  R765  (1999).

\bibitem{cornish1}
S.~L. Cornish {\it et~al.}, e-print cond-mat/0004290 (unpublished).

\end{references}

%
%

\end{document}